\documentstyle[11pt]{article}

\advance \topmargin by -\headheight
\advance \topmargin by -\headsep

\evensidemargin \oddsidemargin
\begin{document}

%%%%%%%%%%%%%%%%%%%%%%%%%%   START OF MACROS  %%%%%%%%%%%%%%%%%%%%%%%%%%
%               MACROS FORMATTING AND EQUATIONS
\def\rf#1{(\ref{eq:#1})}
\def\lab#1{\label{eq:#1}}
\def\nonu{\nonumber}
\def\br{\begin{eqnarray}}
\def\er{\end{eqnarray}}
\def\be{\begin{equation}}
\def\ee{\end{equation}}
\def\eq{\!\!\!\! &=& \!\!\!\! }
\def\foot#1{\footnotemark\footnotetext{#1}}
\def\lb{\lbrack}
\def\rb{\rbrack}
\def\llangle{\left\langle}
\def\rrangle{\right\rangle}
\def\blangle{\Bigl\langle}
\def\brangle{\Bigr\rangle}
\def\llb{\left\lbrack}
\def\rrb{\right\rbrack}
\def\Blb{\Bigl\lbrack}
\def\Brb{\Bigr\rbrack}
\def\lcurl{\left\{}
\def\rcurl{\right\}}
\def\({\left(}
\def\){\right)}
\def\v{\vert}                     %% vertical bars
\def\bv{\bigm\vert}               %%
\def\Bgv{\;\Bigg\vert}            %%
\def\bgv{\bigg\vert}              %%
\def\lskip{\vskip\baselineskip\vskip-\parskip\noindent}
\def\mskp{\par\vskip 0.3cm \par\noindent}
\def\sskp{\par\vskip 0.15cm \par\noindent}
\def\bc{\begin{center}}
\def\ec{\end{center}}
\def\Lbf#1{{\Large {\bf {#1}}}}
\def\lbf#1{{\large {\bf {#1}}}}
%              COMMON PHYSICS SYMBOLS
\def\tr{\mathop{\rm tr}}                  % tr - small trace
\def\Tr{\mathop{\rm Tr}}                  % Tr - big trace
\newcommand\partder[2]{{{\partial {#1}}\over{\partial {#2}}}}
                                                  % partial derivative
\newcommand\partderd[2]{{{\partial^2 {#1}}\over{{\partial {#2}}^2}}}
                                      % second-order partial derivative
\newcommand\partderh[3]{{{\partial^{#3} {#1}}\over{{\partial {#2}}^{#3} }}}
                                      % higher-order partial derivative
\newcommand\partderm[3]{{{\partial^2 {#1}}\over{\partial {#2} \partial {#3} }}}
                                      % mixed second-order partial derivative
\newcommand\funcder[2]{{{\delta {#1}}\over{\delta {#2}}}}
                                                % functional derivative
\newcommand\Bil[2]{\Bigl\langle {#1} \Bigg\vert {#2} \Bigr\rangle}  %% <.|.>
\newcommand\bil[2]{\left\langle {#1} \bigg\vert {#2} \right\rangle} %% <.|.>
\newcommand\me[2]{\left\langle {#1}\right|\left. {#2} \right\rangle} %% <.|.>
\newcommand\sbr[2]{\left\lbrack\,{#1}\, ,\,{#2}\,\right\rbrack} % commutator
\newcommand\Sbr[2]{\Bigl\lbrack\,{#1}\, ,\,{#2}\,\Bigr\rbrack}
% (Large)  commutator
\newcommand\pbr[2]{\{\,{#1}\, ,\,{#2}\,\}}       % Poisson brackets
\newcommand\Pbr[2]{\Bigl\{ \,{#1}\, ,\,{#2}\,\Bigr\}}  % Poisson brackets
% (large)
\newcommand\pbbr[2]{\lcurl\,{#1}\, ,\,{#2}\,\rcurl}  % Poisson brackets
% (left-right)
%              MATH SYMBOLS
\def\a{\alpha}
\def\b{\beta}
\def\c{\chi}
\def\d{\delta}
\def\D{\Delta}
\def\eps{\epsilon}
\def\vareps{\varepsilon}
\def\g{\gamma}
\def\G{\Gamma}
\def\grad{\nabla}
\def\h{{1\over 2}}
\def\l{\lambda}
\def\L{\Lambda}
\def\m{\mu}
\def\n{\nu}
\def\ov{\over}
\def\om{\omega}
\def\O{\Omega}
\def\p{\phi}
\def\P{\Phi}
\def\pa{\partial}
\def\tpa{{\tilde \partial}}
\def\pr{\prime}
\def\ra{\rightarrow}
\def\lra{\longrightarrow}
\def\s{\sigma}
\def\S{\Sigma}
\def\t{\tau}
\def\th{\theta}
\def\Th{\Theta}
\def\z{\zeta}
\def\ti{\tilde}
\def\wti{\widetilde}
\def\ca{{\cal A}}
\def\cb{{\cal B}}
\def\ce{{\cal E}}
\def\cd{{\cal D}}
\def\cF{{\cal F}}
\def\cL{{\cal L}}
\newcommand\sumi[1]{\sum_{#1}^{\infty}}   %% summation till infinity
\newcommand\fourmat[4]{\left(\begin{array}{cc}  %%   2x2 matrix
{#1} & {#2} \\ {#3} & {#4} \end{array} \right)}
\newcommand\twocol[2]{\left(\begin{array}{cc}  %%   2 column
{#1} \\ {#2} \end{array} \right)}
%       FAKE BLACKBOARD BOLD MACROS FOR REALS, COMPLEX, ETC.
\font\numbers=cmss12
%\font\numbers=cmu10 scaled\magstep1
\font\upright=cmu10 scaled\magstep1
\def\stroke{\vrule height8pt width0.4pt depth-0.1pt}
\def\topfleck{\vrule height8pt width0.5pt depth-5.9pt}
\def\botfleck{\vrule height2pt width0.5pt depth0.1pt}
\def\Zmath{\vcenter{\hbox{\numbers\rlap{\rlap{Z}\kern 0.8pt\topfleck}\kern
2.2pt
                   \rlap Z\kern 6pt\botfleck\kern 1pt}}}
\def\Qmath{\vcenter{\hbox{\upright\rlap{\rlap{Q}\kern
                   3.8pt\stroke}\phantom{Q}}}}
\def\Nmath{\vcenter{\hbox{\upright\rlap{I}\kern 1.7pt N}}}
\def\Cmath{\vcenter{\hbox{\upright\rlap{\rlap{C}\kern
                   3.8pt\stroke}\phantom{C}}}}
\def\Rmath{\vcenter{\hbox{\upright\rlap{I}\kern 1.7pt R}}}
\def\IZ{\ifmmode\Zmath\else$\Zmath$\fi}
\def\IQ{\ifmmode\Qmath\else$\Qmath$\fi}
\def\IN{\ifmmode\Nmath\else$\Nmath$\fi}
\def\IC{\ifmmode\Cmath\else$\Cmath$\fi}
\def\IR{\ifmmode\Rmath\else$\Rmath$\fi}
\def\one{\hbox{{1}\kern-.25em\hbox{l}}}
\def\0#1{\relax\ifmmode\mathaccent"7017{#1}%
        \else\accent23#1\relax\fi}
\def\omz{\0 \omega}
% THIS DEFINES "REMARK", "PROPOSITION", "THEOREM" ETC.
\newtheorem{definition}{Definition}%[section]
\newtheorem{proposition}{Proposition}%[section]
\newtheorem{theorem}{Theorem}%[section]
\newtheorem{lemma}{Lemma}%[section]
\newtheorem{corollary}{Corollary}%[section]
\def\proof{\par{\it Proof}. \ignorespaces} \def\endproof{{$\Box$}\par}
\newenvironment{Proof}{\proof}{\endproof}
%%%%%%%%%%%%%%%%%%%%%%%     SPECIAL SYMBOLS   %%%%%%%%%%%%%%%%%%%%%%%
\def\Winf{{\bf W_\infty}}               % Linear W-infinity
\def\Win1{{\bf W_{1+\infty}}}           % Linear W-1+infinity
\def\nWinf{{\bf {\hat W}_\infty}}       % Nonlinear W-infinity
\def\hWinf{{\bf {\hat W}_{\infty}}}        % Hat-W-inf = Nonlinear W-inf
%           PSEUDO-DIFFERENTIAL  OPERATORS
\def\PsDA{\Psi{\cal DO}}
                   % algebra of all pseudo-differential operators
%       MACROS  FOR  OAKPARK
\def\cKP{{\sf cKP}~}
\def\scKP{{\sf scKP}~}
\newcommand\Back{{B\"{a}cklund}~}
\newcommand\DB{{Darboux-B\"{a}cklund}~}
\def\BH{{Burgers-Hopf}~}
\def\tQ{{\widetilde Q}}
\def\tit{{\tilde t}}
\def\hQ{{\widehat Q}}
\def\hb{{\widehat b}}
\def\hR{{\widehat R}}
\def\htt{{\hat t}}
\newcommand{\nit}{\noindent}
\newcommand{\ct}[1]{\cite{#1}}
\newcommand{\bi}[1]{\bibitem{#1}}
%       THIS DEFINES THE JOURNAL CITATIONS
\newcommand\PRL[3]{{\sl Phys. Rev. Lett.} {\bf#1} (#2) #3}
\newcommand\NPB[3]{{\sl Nucl. Phys.} {\bf B#1} (#2) #3}
\newcommand\NPBFS[4]{{\sl Nucl. Phys.} {\bf B#2} [FS#1] (#3) #4}
\newcommand\CMP[3]{{\sl Commun. Math. Phys.} {\bf #1} (#2) #3}
\newcommand\PRD[3]{{\sl Phys. Rev.} {\bf D#1} (#2) #3}
\newcommand\PLA[3]{{\sl Phys. Lett.} {\bf #1A} (#2) #3}
\newcommand\PLB[3]{{\sl Phys. Lett.} {\bf #1B} (#2) #3}
\newcommand\JMP[3]{{\sl J. Math. Phys.} {\bf #1} (#2) #3}
\newcommand\PTP[3]{{\sl Prog. Theor. Phys.} {\bf #1} (#2) #3}
\newcommand\SPTP[3]{{\sl Suppl. Prog. Theor. Phys.} {\bf #1} (#2) #3}
\newcommand\AoP[3]{{\sl Ann. of Phys.} {\bf #1} (#2) #3}
\newcommand\RMP[3]{{\sl Rev. Mod. Phys.} {\bf #1} (#2) #3}
\newcommand\PR[3]{{\sl Phys. Reports} {\bf #1} (#2) #3}
\newcommand\FAP[3]{{\sl Funkt. Anal. Prilozheniya} {\bf #1} (#2) #3}
\newcommand\FAaIA[3]{{\sl Functional Analysis and Its Application} {\bf #1}
(#2) #3}
\newcommand\TAMS[3]{{\sl Trans. Am. Math. Soc.} {\bf #1} (#2) #3}
\newcommand\InvM[3]{{\sl Invent. Math.} {\bf #1} (#2) #3}
\newcommand\AdM[3]{{\sl Advances in Math.} {\bf #1} (#2) #3}
\newcommand\PNAS[3]{{\sl Proc. Natl. Acad. Sci. USA} {\bf #1} (#2) #3}
\newcommand\LMP[3]{{\sl Letters in Math. Phys.} {\bf #1} (#2) #3}
\newcommand\IJMPA[3]{{\sl Int. J. Mod. Phys.} {\bf A#1} (#2) #3}
\newcommand\TMP[3]{{\sl Theor. Mat. Phys.} {\bf #1} (#2) #3}
\newcommand\JPA[3]{{\sl J. Physics} {\bf A#1} (#2) #3}
\newcommand\JSM[3]{{\sl J. Soviet Math.} {\bf #1} (#2) #3}
\newcommand\MPLA[3]{{\sl Mod. Phys. Lett.} {\bf A#1} (#2) #3}
\newcommand\JETP[3]{{\sl Sov. Phys. JETP} {\bf #1} (#2) #3}
\newcommand\JETPL[3]{{\sl  Sov. Phys. JETP Lett.} {\bf #1} (#2) #3}
\newcommand\PHSA[3]{{\sl Physica} {\bf A#1} (#2) #3}
\newcommand\PHSD[3]{{\sl Physica} {\bf D#1} (#2) #3}
\newcommand\JPSJ[3]{{\sl J. Phys. Soc. Jpn.} {\bf #1} (#2) #3}
\newcommand\JGP[3]{{\sl J. Geom. Phys.} {\bf #1} (#2) #3}
%%%%%%%%%%%%%%%%%%%%%%%%%%   END OF MACROS  %%%%%%%%%%%%%%%%%%%%%%%%%%
\begin{titlepage}
\vspace*{-1cm}
\noindent
January, 1996 \hfill{INRNE-TH/96-3}\\
${}$ \hfill{UICHEP-TH/96-08} \\
${}$ \hfill{hep-th/9602068}
\begin{center}
{\Large {\bf Virasoro Symmetry of Constrained KP Hierarchies}}  
\end{center}
\vskip .3in
\begin{center}
{ H. Aratyn\footnotemark
\footnotetext{Work supported in part by U.S. Department of Energy,
contract DE-FG02-84ER40173}}

\par \vskip .1in \noindent
Department of Physics \\
University of Illinois at Chicago\\
845 W. Taylor St.\\
Chicago, IL 60607-7059, {\em e-mail}:
aratyn@uic.edu \\
\par \vskip .3in
E. Nissimov${}^{a),\, 2}$
and S. Pacheva${}^{a),\, b),}$\foot{Supported in part
by Bulgarian NSF grant {\em Ph-401}.}
\par \vskip .1in \noindent
${}^{a)}$ Institute of Nuclear Research and Nuclear Energy \\
Boul. Tsarigradsko Chausee 72, BG-1784 ~Sofia, Bulgaria \\
{\em e-mail}: emil@bgearn.bitnet, svetlana@bgearn.bitnet \\
and \\
${}^{b)}$ Department of Physics, Ben-Gurion University of the Negev \\
Box 653, IL-84105 $\;$Beer Sheva, Israel \\
{\em e-mail}: emil@bgumail.bgu.ac.il, svetlana@bgumail.bgu.ac.il
\end{center}
\vskip .3in

\begin{abstract}
Additional non-isospectral symmetries are formulated for the constrained
Kadomtsev-Petvi\-a\-shvi\-li (\cKP$\!\!$) integrable hierarchies.
The problem of compatibility of additional
symmetries with the underlying constraints is solved explicitly
for the Virasoro part of the additional symmetry
through appropriate modification of the standard additional-symmetry flows for
the general (unconstrained) KP hierarchy.
We also discuss the
special case of \cKP --
truncated KP hierarchies, obtained
as \DB orbits of initial purely differential  Lax operators.
The latter give rise to
Toda-lattice-like structures relevant for discrete
(multi-)matrix models.
Our construction establishes the condition for commutativity of the
additional-symmetry
flows with the discrete \DB transformations
of \cKP hierarchies
leading to a new derivation of the
string-equation constraint in matrix models.

\end{abstract}
\end{titlepage}

\underline{Introduction.}$\;\,$
Relations between integrable models and conformal symmetries
have been studied intensely since the first early signs of
their interconnection
showed up in the literature in seventies \ct{early}.
More recently, the KdV hierarchy formulation of nonperturbative $2$-d quantum
gravity \ct{integr-matrix}
in the framework of (multi-)matrix models
prompted more studies in this field.
The subsequent work pointed out the non-isospectral symmetry origin of the
pertinent
Virasoro constraints
on the string partition function
but remained mostly limited to the KdV-like reduction
of the KP hierarchy since it was dealing with the double scaling limit of
the matrix models \ct{w-kdv}.

Quite recently a new class of integrable systems appeared both in
mathematical literature \ct{oevelr} and independently in physics literature
\ct{multi-b}, where the motivation came from Toda field theory and discrete
matrix models.
These systems belong to
class of the
 so called constrained KP hierarchies (\cKP)
as they are obtained by a symmetry reduction (which generalizes the KdV
type of reduction) from the underlying
general (unconstrained)
KP hierarchy.
\cKP hierarchies contain a large number of interesting hierarchies of
soliton equations.

We address here the issue of formulation of additional
non-isospectral
Virasoro symmetry structure for the \cKP hierarchies, or in different words,
solving the problem for
compatibility of the constraints with the additional
non-isospectral
symmetries of the original KP hierarchy. It is shown how initially
the Virasoro algebra is broken by the constraints down
to the $sl(2)$ subalgebra (containing Galilean and scaling symmetries) of the
Virasoro algebra and how the Virasoro symmetry (for the Virasoro generators
$\cL_n\, , \, n\geq -1$) is recovered by modifying generators via
adding a new structure consisting of ghost flows related to the plethora of
(adjoint) eigenfunctions characteristic for the \cKP hierarchies.

We also discuss
the special case of \cKP hierarchies --
the so called truncated KP hierarchies obtained
as \DB (DB) orbits of initial purely differential  Lax operators.
They are associated with Toda-lattice-like discrete integrable systems which
can be naturally embedded in the \cKP hierarchies and
which are relevant for the description of
discrete (multi-)matrix models.
Applying here our construction establishes the condition for commutativity of
additional-symmetry
flows with the discrete \DB transformations.
This condition sheds new light on the derivation of the
string-equation constraint (string condition)
for matrix models. Details of calculations will appear elsewhere \ct{oakpark}.

\underline{Background on KP Hierarchy.}$\;\,$
We use the calculus of the pseudodifferential operators to describe the
KP hierarchy of nonlinear evolution equations.
In what follows the operator $D$ is such that $ \sbr{D}{f} = f^{\pr}$
with $f^{\pr}= \pa f = \pa f /\pa x$ and it satisfies the
generalized Leibniz rule
(eq.\rf{gleib} from Appendix).

The main object here is a pseudo-differential Lax operator $L$
of a generalized KP hierarchy:
\be
L = D^r + \sum_{j=0}^{r-2} v_j D^j + \sum_{i \geq 1} u_i D^{-i}
\lab{gen-KP}
\ee
The associated Lax equations
\be
\partder{}{t_l} L = \Sbr{L^{l\over r}_{+}}{L}
\quad    \; \; l = 1, 2, \ldots
\lab{lax-eq}
\ee
(recall that $x \equiv t_1$) describe isospectral deformations of $L$.
In \rf{lax-eq}and below, the subscripts $(\pm )$ of pseudo-differential 
operators indicate purely differential/pseudo-differential parts.
Commutativity of the isospectral flows
$\partder{}{t_l}$ \rf{lax-eq}
is then assured by the Zakharov-Shabat equations.
One can also represent the Lax operator in terms of the dressing
operator $W= 1 + \sum_1^{\infty} w_n D^{-n}$ through
$L= W \, D^r \,W^{-1}$.
In this framework equation \rf{lax-eq} is equivalent to the so called
Wilson-Sato equation:
\be
\partder{}{t_l} W = - \( W D^l W^{-1} \)_{-} W
\lab{sato-a}
\ee

For a given Lax  operator $L$, which satisfies Sato's flow equation
\rf{lax-eq}, we call the function $\Phi$ ($ \Psi$), whose flows are
given by the expression\foot{For any (pseudo-)differential operator
$A$ and a function $f$, the symbol $A(f)$ will indicate application
(action) of $A$ on $f$ as opposed to the symbol $Af$ meaning just operator
product of $A$ with the zero-order (multiplication) operator $f$.} :
\be
\partder{\Phi}{t_l} = L^{l \ov r}_{+} (\Phi ) \qquad; \qquad
\partder{\Psi}{t_l} = - \( L^{\ast} \)^{l \ov r}_{+} (\Psi )
\quad\;\; l=1,2, \ldots
\lab{eigenlax}
\ee
an {\it (adjoint) eigenfunction} of $L$. In \rf{eigenlax} we have
introduced an operation of conjugation, defined by simple rules
$D^{\ast} = -D$ and $(AB)^{\ast}= B^{\ast} A^{\ast}$.
An eigenfunction, which in addition also satisfies
the spectral equations $L\psi (\l ,t) = \l\psi (\l ,t)$ is called
{\em Baker-Akhiezer (BA) function}.

\underline{Additional Symmetries for the KP hierarchy.}$\;\,$
The KP hierarchy has an infinite set of commuting symmetries associated
with the isospectral flows described above
(eq.\rf{lax-eq}).
However, the group of symmetries of the KP hierarchy is known to be much
bigger. The extra symmetries are called ``non-isospectral'' or
``additional'' symmetries.
A convenient approach to deal with symmetries of the integrable
hierarchies of equations was developed by Orlov and Schulman
(see \ct{add-symm,cortona,moerbeke}) and this is the approach we will use in
this paper.
Other important contributions to the subject of additional symmetries
for the KP hierarchy were made by Fuchssteiner \ct{Fuchs} and Chen et al.
\ct{Chenetal}. See also references \ct{lie-akns} for the related discussion
of the AKNS model, \ct{aoyama} for the truncated KP hierarchy and \ct{HMG94}
for treatment of the generalized matrix hierarchies.

Let $M$ be  an operator ``canonically conjugated''
to $L$ such that:
\be
\Sbr{L}{M} = \one \quad , \quad
\partder{}{t_l} M = \Sbr{L^{l\over r}_{+}}{M}
\lab{L-M}
\ee
The $M$-operator can be expressed in terms of dressing of the
``bare'' $M^{(0)}$ operator
\be
M^{(0)} = \sum_{l \geq 1} \frac{l}{r} t_l D^{l-r} =
X_{(r)} + \sum_{l \geq 1} \frac{l+r}{r} t_{r+l} D^l \quad ; \quad
X_{(r)} \equiv \sum_{l=1}^{r} \frac{l}{r} t_l D^{l-r}
\lab{M-0}
\ee
conjugated to the ``bare'' Lax operator $L^{(0)} = D^r$. The dressing gives
\br
M \eq  W M^{(0)} W^{-1} =
W X_{(r)} W^{-1} + \sum_{l \geq 1} \frac{l+r}{r} t_{r+l} L^{l\over r} =
\sum_{l \geq 0} \frac{l+r}{r} t_{r+l} L^{l\over r}_{+} + M_{-}
\lab{M-dress}  \\
M_{-} \eq W X_{(r)} W^{-1} - t_r -
\sum_{l \geq 1} \frac{l+r}{r} t_{r+l} \partder{W}{t_l}  \, .\, W^{-1}
\lab{M--}
\er
where in \rf{M--} we used eqs.\rf{sato-a}.
Note that $X_{(r)}$ is a pseudo-differential operator satisfying
$\Sbr{D^r}{X_{(r)}} = \one$ .

The so called {\em additional (non-isospectral) symmetries}
\ct{add-symm,cortona} are defined as vector fields on the space of
${\bf KP}$ Lax operators \rf{gen-KP} or, alternatively, on the dressing
operator through their flows as follows:
\be
{\bar \pa}_{k,n} L = - \Sbr{\( M^n L^k\)_{-}}{L} =
\Sbr{\( M^n L^k\)_{+}}{L} + n M^{n-1} L^k  \;\,;  \;\;\;\,
{\bar \pa}_{k,n} W = - \( M^n L^k\)_{-} W
\lab{add-symm-L}
\ee
The additional flows commute with the usual KP hierarchy flows given
in \rf{lax-eq}. But they do not commute among themselves,
instead they form the $\Win1$ algebra (see e.g. \ct{cortona}).
One finds that the Lie algebra of operators ${\bar \pa}_{k,n}$
is isomorphic to the Lie algebra generated by $- z^n (\pa/\pa z)^k$.
Especially for $n=1$ this becomes an isomorphism to the Virasoro
algebra ${\bar \pa}_{k,1} \sim - \cL_{k-1}$, with $\sbr{\cL_n}{\cL_k} = (n-k)
\cL_k$.

\underbar{Constrained KP Hierarchy and Additional Symmetry.}$\;\,$
We now turn to the main problem of this letter, namely,
compatibility of the additional Virasoro symmetry
with the constraints defining the \cKP hierarchy.
We first introduce the symmetry constraints leading to the \cKP hierarchy.
Let $\pa_{\a_{i}}$ be vector fields, whose action on the
standard KP
Lax operator $Q=D + \sumi{i=0} u_i D^{-i-1} $ is induced by
the (adjoint) eigenfunctions $\P_i, \Psi_i$ of $Q$ through \ct{oevelr}:
\be
\pa_{\a_{i}} {Q} \equiv \lb Q \,  , \, \Phi_i D^{-1} \Psi_i \rb
\lab{ghostflo}
\ee
We have the following proposition:
\begin{proposition}
The vector fields $\pa_{{\a_{i}}}$ commute with the isospectral flows of the
Lax operator $Q$:
\be
\sbr{\pa_{\a_{i}}}{\partder{}{t_l}} Q = 0  \qquad l=1, 2, \ldots
\lab{comm}
\ee
\label{proposition:commflo}
\end{proposition}
\begin{definition}
The constrained KP hierarchy (denoted as ${\sf cKP}_{r,m}$) is obtained by
identifying the ``ghost'' flow $\sum_{i=1}^m \pa_{{\a_{i}}}$  with the
isospectral flow $\pa_r$ of the
original
KP hierarchy.
\label{definition:ckpdef}
\end{definition}
Comparing \rf{ghostflo} with equation \rf{lax-eq} we find that for the Lax
operator belonging to the ${\sf cKP}_{r,m}$ hierarchy we have
$Q^r_{-} = \sum_{i=1}^m \Phi_i D^{-1} \Psi_i $.
Hence we are led to the Lax operator $L = Q^r$ given by
\be
L = L_{+} +  \sum_{i=1}^m \Phi_i D^{-1} \Psi_i
=D^r+ \sum_{l=0}^{r-2} v_l D^l + \sum_{i=1}^m \Phi_i D^{-1} \Psi_i
\lab{f-5}
\ee
and subject to the Lax equation \rf{lax-eq}.
Therefore, we parametrize the ${\sf cKP}_{r,m}$ hierarchy
in terms of the Lax operator \rf{f-5}. Note that the
(adjoint) eigenfunctions $\Phi_i$, $ \Psi_i$ of the original Lax operator
$Q$
used in the above construction remain
(adjoint) eigenfunctions for $L$  \rf{f-5} \ct{avoda}.

Applying
the additional-symmetry flows \rf{add-symm-L} on $L$ \rf{f-5}
for $n=1$ we get
\be
\( {\bar \pa}_{k,1} L\)_{-} =
{\sbr{\( M L^k \)_{+}}{L}}_{-} + \( L^k  \)_{-}
\lab{symm1}
\ee
Using the simple identities \rf{tkppsi} and \rf{lkminus}
from Appendix
for the Lax operator \rf{f-5}, we are able to rewrite \rf{symm1} as:
\be
\( {\bar \pa}_{k,1} L\)_{-} = \sum_{i=1}^m \( M L^k \)_{+} (\Phi_i) D^{-1}
\Psi_i - \sum_{i=1}^m \Phi_i D^{-1} \( M L^k \)^{\ast}_{+} ( \Psi_i )  +
\sum_{i=1}^m \sum_{j=0}^{k-1} L^{k-j-1} (\Phi_i) D^{-1}
\( L^{\ast}\)^{j} ( \Psi_i )
\lab{pak1}
\ee
Here
\be
L (\Phi_i ) \equiv
L_{+} (\Phi_i ) + \sum_{j=1}^m \Phi_j \pa_x^{-1} \( \Psi_j \Phi_i\)
\lab{applic}
\ee
(and similarly for the adjoint counterpart) denotes action of $L$ on
$\Phi_i$.  
Notice that $L^{k-j-1} (\Phi_i) $,  $\( L^{\ast}\)^{j} ( \Psi_i) $ are 
(adjoint) eigenfunctions of $L$ \rf{f-5}. Hence, 
whereas the original $L$ \rf{f-5} belongs to the class of ${\sf cKP}_{r,m}$
hierarchies, the transformed Lax operator given by
${\bar \pa}_{k,1} L$ (cf. eq.\rf{pak1}) belongs to a {\em different} class --
${\sf cKP}_{r,m(k-1)}$ (for $k \geq 3$), since the number of eigenfunctions
has increased.

For $k=0,1,2$ the flow equations \rf{pak1} can still
be rewritten in the desired original ${\sf cKP}_{r,m}$ form:  %%EA
\be
\(\pa_{\t} L\)_{-} = \sum_{i=1}^m \(\pa_{\t} \Phi_i\) D^{-1} \Psi_i + \Phi_i
D^{-1} \( \pa_{\t} \Psi_i\)
\lab{addflo}
\ee
with
$\pa_\t \equiv {\bar \pa}_{k,1}\; (k=0,1,2)$, where:
\br
{\bar \pa}_{0,1} \Phi_i \eq \( M\)_{+}  (\Phi_i) \qquad;\qquad
{\bar \pa}_{0,1} \Psi_i = - \( M\)^{\ast}_{+} ( \Psi_i  ) \lab{papsiz} \\
{\bar \pa}_{1,1} \Phi_i \eq \( ML\)_{+} (\Phi_i) + \a \Phi_i
\quad;\quad
{\bar \pa}_{1,1} \Psi_i = - \( ML\)^{\ast}_{+} (\Psi_i) + \b \Psi_i
\quad \; \a+\b =1\lab{papsi1} \\
{\bar \pa}_{2,1} \Phi_i \eq \( M L^2 \)_{+} (\Phi_i) + L (\Phi_i)
\quad;\quad
{\bar \pa}_{2,1} \Psi_i = - \( M L^2 \)^{\ast}_{+} (\Psi_i) + L^{\ast}
(\Psi_i)
\lab{papsi2}
\er
Note an ambiguity on the right hand sides of \rf{papsi1}.

The fact, that the action of the
additional-symmetry
flows can be defined on the (adjoint) eigenfunctions
$\Phi_i ,\Psi_i\,$ (as in \rf{papsiz}--\rf{papsi2}) in a way which is
consistent with the additional-symmetry
flows \rf{symm1} on the constrained Lax operator \rf{f-5},
is equivalent to compatibility of the constraints
with additional-symmetry flows for the KP hierarchy.

Since the additional flows satisfy an algebra
$\sbr{{\bar \pa}_{l,1}}{{\bar \pa}_{k,1}} = -\( l-k \) {\bar \pa}_{l+k-1,1}$
we have an isomorphism ${\bar \pa}_{k,1} \sim - \cL_{k-1}$ with the Virasoro
operators and equations \rf{papsiz}--\rf{papsi2} contain the
$sl(2) $ subalgebra
generators $\cL_{-1}, \cL_0 , \cL_1$.

Since for $\pa_\t \equiv {\bar \pa}_{k,1}\; k \geq 3$, eq.\rf{addflo} does not
hold anymore due to absence of consistent definitions for
${\bar \pa}_{k,1} \Phi_i ,\, {\bar \pa}_{k,1} \Psi_i$ generalizing
\rf{papsiz}--\rf{papsi2} for higher $k$, it appears
that the symmetry constrains behind the \cKP hierarchies have
broken the additional Virasoro symmetry down to the $sl(2)$ subalgebra.

To recover the complete Virasoro symmetry,
our strategy will be to redefine the additional-symmetry
generators.
We first describe our technique for $k=3$ in which case
equation \rf{symm1} contains a term:
\be
\( L^3 \)_{-} = \sum_{i=1}^m \Phi_i D^{-1} \(L^{\ast}\)^2 \( \Psi_i\)  +
\sum_{i=1}^m  L\(\Phi_i\) D^{-1} L^{\ast} \( \Psi_i\)  +
\sum_{i=1}^m  L^2\(\Phi_i\) D^{-1} \Psi_i
\lab{l3m}
\ee
Note that the middle term in \rf{l3m} is not of the form of required
by equation \rf{addflo}.
At this point we recall that for
\be
X \equiv \sum_{k=1}^I M_k D^{-1} N_k
\lab{xdef}
\ee
with definitions \rf{f-5} and \rf{xdef} we find using identity \rf{x1x2} from
Appendix:
\be
{\sbr{X}{L}}_{-} = \sum_{k=1}^I \( - L (M_k) D^{-1} N_k + M_k D^{-1} L^{\ast}
(N_k)\) + \sum_{i=1}^m \(  X ( \Phi_i ) D^{-1} \Psi_i
- \Phi_i D^{-1} X^{\ast} (\Psi_i )  \)
\lab{xlbra}
\ee
According to \ref{proposition:commflo}
the flows generated by \rf{xlbra} will commute with the isospectral flows
\rf{lax-eq}
provided
$M_i,N_i$ are (adjoint) eigenfunctions, which will be the case in what follows.
Considering now as an example:
\be
Y_3 \equiv \h \sum_{i=1}^m  \Bigl( \Phi_i D^{-1} L^{\ast} \( \Psi_i\) -
L \( \Phi_i\) D^{-1}  \Psi_i \Bigr)
\lab{xdef3}
\ee
we find that
\br
{\sbr{Y_3}{L}}_{-} \eq -\( L^3 \)_{-} + {3 \over 2} \sum_{i=1}^m
\( \Phi_i D^{-1} \(L^{\ast}\)^2 \( \Psi_i\)  + L^2\(\Phi_i\) D^{-1} \Psi_i \)
\nonu \\
&+& \sum_{i=1}^m \Bigl( Y_3 ( \Phi_i )  D^{-1} \Psi_i
- \Phi_i D^{-1} Y^{\ast}_3 (\Psi_i )  \Bigr)
\lab{l3xl}
\er
Hence ${\sbr{-\( ML^3 \)_{-}+Y_3}{L}}_{-}$ still has a form of
\rf{addflo}.
Therefore, it may be possible to find additional symmetries for the \cKP model
by combining
the original
${\bar \pa}_{k,1}$ flows
and the so-called ghost flows (associated with operators of $Y_3$ type).
This will work provided that the above construction  
yields the Virasoro generator  $\cL_2$  obeying
the correct algebra with the unbroken
$sl(2)$ generators found above in \rf{papsiz}--\rf{papsi2}.
Note that each of the two terms in \rf{xdef3} could have been used with an
appropriate factor to obtain the similar conclusion as in \rf{l3xl}.
The choice of
the coefficients in $Y_3$  \rf{xdef3},
apart from being the most symmetric
combination, has the advantage that it will lead below to the correct
Virasoro algebra commutators.

We now generalize the above manipulations to an arbitrary $k$.
We introduce
the pseudo-differential operators:
\be
Y_k  \equiv   \sum_{i=1}^m \sum_{j=0}^{k-2} \(j - \h (k-2)\) L^{k-2-j}
(\Phi_i) D^{-1} \( L^{\ast}\)^{j} (\Psi_i) \quad ;\quad k \geq 2
\lab{xkb}
\ee
which are non-zero for $k \geq 3$.
Notice that \rf{xlbra} and
identity \rf{lkminus} from Appendix
enable us to obtain:
\br
{\sbr{Y_k}{L}}_{-} &=& {k \over 2} \sum_{i=1}^m \( \Phi_i D^{-1}
\( L^{\ast}\)^{k-1} (\Psi_i) +  L^{k-1} (\Phi_i) D^{-1} \Psi_i\)
\lab{lkminnn} \\
&-&\(  L^k \)_{-} + \sum_{i=1}^{m} \Bigl( -\Phi_i D^{-1}
Y_k^{\ast} (\Psi_i) + Y_k (\Phi_i) D^{-1} \Psi_i \Bigr)
\nonu
\er

Our main result is contained in the following
\begin{proposition}
The correct additional-symmetry flows for the \cKP hierarchies \rf{f-5},
spanning the Virasoro algebra, are given by:
\be
\pa^{\ast}_k\, L  \;\equiv\;  \sbr{- \( M L^k \)_{-} + Y_k}{L}
\lab{pasta}
\ee
{\sl i.e.}, with the following isomorphism
$\cL_{k-1} \sim -\( M L^k \)_{-}  + Y_k$, where $Y_k$ are defined in \rf{xkb}.
\label{proposition:mainprop}
\end{proposition}

Indeed, using \rf{lkminnn} we first
find that $\(\pa^{\ast}_k L\)_{-}$ can be
cast in the form of \rf{addflo} with
\be
\pa^{\ast}_k\, \Phi_i =\( M L^k \)_{+} (\Phi_i) + {k \over 2} L^{k-1} (\Phi_i)
+Y_k (\Phi_i)
\;\, ; \; \,
\pa^{\ast}_k\, \Psi_i = - \( M L^k \)^{\ast}_{+} (\Psi_i) + {k \over 2}
(L^{\ast})^{k-1} (\Psi_i) - Y^{\ast}_k (\Psi_i)
\lab{paast}
\ee
Taking into account that $Y_i =0$ for $i=0,1,2$ we see that eq.\rf{paast}
reproduces
\rf{papsiz}-\rf{papsi2} (with ambiguity on the right hand side of \rf{papsi1}
removed by fixing $\a=\b=1/2$).
Hence $\pa^{\ast}_{\ell} = {\bar \pa}_{\ell,1}$ for $\ell = 0,1,2$.

Secondly, we note that the modified additional flows defined by \rf{pasta}
commute with the isospectral flows \rf{lax-eq} and, due to \rf{paast}, they
preserve the form of the \cKP Lax operator \rf{f-5}.
The remaining question is whether they form a closed algebra.
To answer this question we first calculate
\br
\pa^{\ast}_{\ell}\, L^k (\Phi_i) \eq \( M L^\ell \)_{+} \(L^k (\Phi_i)\) +
\( k+ \h \ell \) L^{k+\ell-1} (\Phi_i) \nonu \\
\pa^{\ast}_\ell (L^{\ast})^{k} (\Psi_i) \eq - \( M L^\ell \)^{\ast}_{+}
\((L^{\ast})^{k}\,(\Psi_i)\) + \( k + \h \ell \)
(L^{\ast})^{k+\ell-1} (\Psi_i)
\lab{usepaast}
\er
valid for $\ell = 0,1,2$ and $k \geq 0$, and use it to obtain the following
identity:
\be
{\bar \pa}_{\ell,1}\, Y_k \,= \, \pa^{\ast}_{\ell}\, Y_k \,= \,
{\sbr{\( M L^{\ell}\)_{+}}{Y_k}}_{-} + \( k-\ell \) Y_{k+\ell-1} \quad \;
; \quad \; \ell= 0,1,2
\lab{paiyk}
\ee
Accordingly, for $\ell = 0,1,2$ and any $k \geq 0$ we arrive at the fundamental
commutation relations:
\be
\llb\, \pa^{\ast}_{\ell}\, , \,   \pa^{\ast}_k\, \rrb\, L\,= \,
\( k-\ell \) \, \pa^{\ast}_{k+\ell-1}\, L
\lab{paipay}
\ee
The above discussion shows that
$\sbr{\cL_i}{\cL_{k}} = (i-k) \cL_{i+k}$ for $i=-1,0,1$ ($sl(2)$ generators)
and arbitrary $k$, where $\cL_{k-1} \sim - \pa^{\ast}_{k}$.
Accordingly, if $\cL_2$ is associated with $Y_3 - \( M L^3 \)_{-}$
all higher Virasoro operators can be obtained recursively from
\be
\cL_{n+1} = { -1 \over (n-1)} \sbr{\cL_{n}}{\cL_{1}}
\lab{recursa}
\ee
One can now easily show by induction that $\cL_k\, , \, k\geq -1$ obtained
in the above way form a closed Virasoro
algebra up to terms, which commute with the $sl(2)$ generators.
For illustration consider, {\sl e.g.}, $\sbr{\cL_2}{\cL_3}\equiv Z $. Commuting
$\cL_{-1}$ with $Z$ we find $\sbr{\cL_{-1}}{Z} = 6 \cL_4$, which
fixes $Z$ to be $-\cL_5$ up to terms commuting with the $sl(2)$ subalgebra.
It is easy to see how to extend this argument to cover the whole Virasoro
algebra.

\underline{\DB of \cKP Hierarchies. Truncated KP Hierarchies.}$\;\,$
Let $\Phi$ be an eigenfunction of $L$ defining a \DB transformation,
{\sl i.e.} :
\be
\partder{}{t_l} \Phi = L^{l\over r}_{+} (\Phi )  \quad ,\quad
{\wti L} = \(\Phi D \Phi^{-1}\)\, L\, \(\Phi D^{-1} \Phi^{-1}\)
\quad ,\quad
{\wti W} =  \(\Phi D \Phi^{-1}\)\, W\, D^{-1}
\lab{gen-L-DB}
\ee
Then the DB-transformed $M$ operator (cf. \rf{M-dress}) acquires the form:
\br
{\wti M} \eq \(\Phi D \Phi^{-1}\)\, M\, \(\Phi D^{-1} \Phi^{-1}\) =
\sum_{l \geq 0} \frac{l+r}{r} t_{r+l} {\wti L}^{l\over r}_{+} + {\wti M}_{-}
\lab{gen-M-DB}  \\
{\wti M}_{-} \eq {\wti W} {\wti X}_{(r)} {\wti W}^{-1} - t_r -
\sum_{l \geq 1} \frac{l+r}{r} t_{r+l}\partder{}{t_l}{\wti W}\,.\,{\wti W}^{-1}
\lab{M--ti}
\er
where ${\wti X}_{(r)} = D X_{(r)} D^{-1}$ with $X_{(r)}$ as in \rf{M-0}.
Clearly ${\wti X}_{(r)}$, like $X_{(r)}$, is
also admissible as canonically conjugated to $D^r$.

In particular, for $L$ belonging to a \cKP hierarchy \rf{f-5} we consider
special class of DB transformations \rf{gen-L-DB} which preserve the
constrained \cKP form of $L$ :
\br
{\wti L} \eq  T_a L T_a^{-1} =
{\wti L}_{+} + \sum_{i=1}^m {\wti \Phi}_i D^{-1} {\wti \Psi}_i
\qquad , \qquad T_a \equiv \Phi_a D \Phi_a^{-1}
\lab{cKP-L-DB} \\
{\wti \Phi}_a \eq T_a L (\Phi_a ) \qquad ,\qquad {\wti \Psi}_a = \Phi_a^{-1}
\lab{DB-1} \\
{\wti \Phi}_i \eq  T_a (\Phi_i ) \quad ,\quad
{\wti \Psi}_i = {T_a^{-1}}^\ast \Psi_i =
- \Phi_a^{-1} \pa_x^{-1} ( \Psi_i \Phi_a ) \qquad, \quad i \neq a
\lab{DB-2}
\er
where the DB-generating $\Phi \equiv \Phi_a$ coincides with one of the
eigenfunctions of the initial $L$ \rf{f-5}.

Let us consider the following generic class of DB orbits on
${\sf cKP}_{r,m}$ :
within each subset of $m$ successive DB steps we
perform DB transformations \rf{cKP-L-DB} w.r.t. the $m$
different eigenfunctions of \rf{f-5}. Repeated use of a
composition formula for Wronskians (see eqs.\rf{iw}--\rf{W-def} from Appendix)
leads us to the following explicit expressions for the eigenfunctions and the
$\t$-function after $km+l$ steps ($1 \leq l \leq m$) of successive DB
transformations (see also ref.\ct{pirin}) :
\br
\Phi_i^{(km+l)} = T^{(km+l-1)}_l \cdots T^{(km)}_1
T^{(km-1)}_m \cdots T^{((k-1)m)}_1 \cdots T^{(m-1)}_m \cdots T^{(0)}_1
\chi^{(k_{\pm})}_i   \nonu  \\
= \frac{W\llb \Phi^{(0)}_1 ,\ldots ,\Phi^{(0)}_m ,
\chi^{(1)}_1,\ldots ,\chi^{(1)}_m ,\ldots ,\chi^{(k-1)}_1,\ldots ,
\chi^{(k-1)}_m , \chi^{(k)}_1,\ldots ,\chi^{(k)}_l , \chi^{(k_{\pm})}_i
\rrb}{W\llb \Phi^{(0)}_1 ,\ldots ,\Phi^{(0)}_m ,
\chi^{(1)}_1,\ldots ,\chi^{(1)}_m ,\ldots ,\chi^{(k-1)}_1,\ldots ,
\chi^{(k-1)}_m ,\chi^{(k)}_1,\ldots ,\chi^{(k)}_l \rrb}
\lab{pchi-a-1}  \\
\chi^{(k_{+})}_i \equiv \chi^{(k+1)}_i \quad {\rm for} \;\; 1 \leq i \leq l
\qquad ; \qquad
\chi^{(k_{-})}_i \equiv \chi^{(k)}_i \quad {\rm for} \;\; l+1 \leq i \leq m
\nonu  \\
\frac{\t^{(km+l)}}{\t^{(0)}} = \Phi^{(km+l-1)}_l \cdots \Phi^{(km)}_1
\Phi^{(km-1)}_m \cdots \Phi^{((k-1)m)}_1 \cdots
\Phi^{(m-1)}_m \cdots \Phi^{(0)}_1
\nonu  \\
= W\llb \Phi^{(0)}_1 ,\ldots ,\Phi^{(0)}_m ,
\chi^{(1)}_1,\ldots ,\chi^{(1)}_m,\ldots ,\chi^{(k-1)}_1,\ldots ,\chi^{(k-1)}_m
 ,
\chi^{(k)}_1,\ldots ,\chi^{(k)}_l \rrb
\lab{tauok-1}
\er
where the upper indices in brackets indicate the order of the corresponding
DB step, the zero index referring to the ``initial'' \cKP Lax operator,
and where we have employed the short-hand notations:
\be
T^{(k)}_i \equiv \Phi^{(k)}_i D \(\Phi^{(k)}_i\)^{-1} \qquad ;\qquad
\chi^{(s)}_i \equiv \( L^{(0)}\)^s (\Phi^{(0)}_i ) \quad ,\;\; i=1,\ldots ,m
\lab{defchi-i}
\ee

As seen from \rf{cKP-L-DB}--\rf{DB-2} and \rf{pchi-a-1}, the DB orbit
$L^{(k)} = \( L^{(k)}\)_{+} + \sum_{i=1}^m \Phi^{(k)}_i D^{-1} \Psi^{(k)}_i$
of ${\sf cKP}_{r,m}$, starting from a purely differential initial
$L^{(0)} = \( L^{(0)}\)_{+}$, defines a class of {\em truncated}
${\sf cKP}_{r,m}$ hierarchies where the $m$ adjoint eigenfunctions
$\Psi_i \equiv \Psi^{(k)}_i$ are not independent from the $m$ eigenfunctions
$\Phi_i \equiv \Phi^{(k)}_i$ since both are parametrized in terms of
$m$ independent functions $\Phi^{(0)}_i$ only.

As a simple example of truncated \cKP hierarchies, consider formulas
\rf{pchi-a-1}--\rf{tauok-1} for the DB orbit of
${\sf cKP}_{1,m}$ hierarchy (the so called ``multi-boson'' reduction of the
general KP hierarchy) starting from a ``free'' initial $L^{(0)} = D$. In
this case we have to substitute in \rf{pchi-a-1}--\rf{tauok-1} :
\be
\chi^{(s)}_i \equiv \pa^s \Phi_i^{(0)} \qquad , \quad
\Phi_i^{(0)} = \int_{\Gamma} d\l \, \phi_i^{(0)} (\l )
\exp \lcurl \sum_{r\geq 1} \l^r t_r \rcurl
\lab{P-0}
\ee
with arbitrary ``densities'' $\phi_i^{(0)} (\l )$ (and with appropriate
contour $\Gamma$ such that the $\l$-integrals exist). A special feature of
truncated ${\sl cKP}_{1,m}$ is that
their dressing operators are truncated (having only finite
number of terms in the pseudo-differential expansion, cf. \ct{aoyama}) :
\be
W^{(km+l)} =  T^{(km+l-1)}_l \cdots T^{(km)}_1 T^{(km-1)}_m
\cdots T^{((k-1)m)}_1 \cdots T^{(m-1)}_m \cdots T^{(0)}_1 D^{-km-l} =
\sum_{j=0}^{km+l} w_j^{(km+l)} D^{-j}
\lab{W-trunc}
\ee
where notations \rf{defchi-i} were used.

The particular case $m=1$ of \rf{cKP-L-DB}--\rf{DB-2},
\rf{pchi-a-1}--\rf{tauok-1} yields:
\br
L^{(k+1)} \eq \(\Phi^{(k)}  D {\Phi^{(k)} }^{-1}\)  \; L^{(k)}
 \; \(\Phi^{(k)}  D^{-1} {\Phi^{(k)} }^{-1}\)
= D + \Phi^{(k+1)} D^{-1} \Psi^{(k+1)}
\lab{lkplus} \\
\Phi^{(k+1)} \eq \Phi^{(k)} \( \ln \Phi^{(k)}\)^{\pr \pr} + \(\Phi^{(k)}\)^2
\Psi^{(k)} \quad ,\quad \Psi^{(k+1)} = \(\Phi^{(k)}\)^{-1}
\lab{pkplus}  \\
\Phi^{(n)} \eq { W_{n+1} \lb \p , \pa \p, \ldots , \pa^n \p \rb \over
W_n \lb \p , \pa \p, \ldots , \pa^{n-1} \p \rb}  \quad , \quad
\t^{(n)} = W_n \lb \p , \pa \p, \ldots , \pa^{n-1} \p \rb
\lab{phik}
\er
where
\be
\p \equiv \Phi^{(0)} = \int d \l \p^{(0)} (\l )
\exp \lcurl \sumi{r=1} t_r \l^r \rcurl
\lab{phi0}
\ee
The hierarchies given by \rf{lkplus} are generalizations of the \BH hierarchy
defined by $L^{(1)}= D + \p ( \ln \p )^{\pr\pr} D^{-1} \p^{-1}$.

\underline{Additional Symmetries {\sl versus} DB Transformations for
\cKP Hierarchies. String Condition.} \\
With the help of identities \rf{DB-like-1}--\rf{DB-like-4} from Appendix
we find the following explicit form of the DB
transformation of the operators $Y_k$ \rf{xkb} :
\br
T_a Y_k T_a^{-1} &=& {\wti Y}_k - \( {\wti L}^{(a)}\)_{-}^{k-1} +
\lcurl T_a \( Y_k + {k\over 2} L^{k-1}\) (\Phi_a )\rcurl D^{-1} \Phi_a^{-1}
\lab{DB-Yk}   \\
\( {\wti L}^{(a)}\)_{-}^{k-1} &\equiv&
\sum_{j=0}^{k-2} {\wti L}^{k-j-2} (\Phi_a) D^{-1}
\( {\wti L}^{\ast}\)^{j} (\Psi_a )
\lab{lkminus-ti-a}
\er
Here ${\wti L},\, T_a $ are as in \rf{cKP-L-DB}
and the DB-transformed ${\wti Y}_k$ have the same form as $Y_k$ in \rf{xkb}
with all (adjoint) eigenfunctions substituted with their
DB-transformed counterparts as in \rf{cKP-L-DB}--\rf{DB-2}. Also notice that
in the particular case of ${\sf cKP}_{r,1}$ hierarchies
$\( {\wti L}^{(a)}\)_{-}^{k-1}$ \rf{lkminus-ti-a} coincides with the
(pseudo-differential part of the power of the) full ${\sf cKP}_{r,1}$ Lax
operator (cf. eq.\rf{f-5} for $m=1$ and \rf{lkminus}).

Taking into account \rf{cKP-L-DB}--\rf{DB-2} and \rf{DB-Yk}--\rf{lkminus-ti-a}
we arrive at the following important
\begin{proposition}
The additional-symmetry flows \rf{pasta} for ${\sf cKP}_{r,1}$ hierarchies
(eq.\rf{f-5} with $m=1$)
commute with the \DB transformations \rf{cKP-L-DB} preserving the form of
${\sf cKP}_{r,1}$, up to shifting of \rf{pasta} by ordinary isospectral flows.
Explicitly we have:
\be
\pa^\ast_k {\wti L} =
- \Sbr{\( {\wti M} {\wti L}^k \)_{-} - {\wti Y}_k}{{\wti L}} +
\partder{\wti L}{t_{k-1}}
\lab{AS-DB-compat}
\ee
\label{proposition:dbcommute}
\end{proposition}

Proposition \ref{proposition:dbcommute} shows that the additional-symmetry
flows \rf{pasta} are well-defined for all ${\sf cKP}_{r,1}$ Lax operators
belonging to a given DB orbit of successive DB transformations. %%BA
Notice that it is precisely the class of (truncated) ${\sf cKP}_{r,1}$
hierarchies which is relevant for the description of discrete (multi-)matrix
models \ct{office,avoda,pirin}.  %%EA

Motivated by applications to (multi-)matrix models (see ref.\ct{oakpark}),
one can require invariance of \cKP hierarchies
under some of the additional-symmetry flows,
{\sl e.g.}, under the lowest one $\pa^\ast_0 \equiv{\bar \pa}_{0,1}$ known as
``string-equation'' constraint (string condition) in the context of the
(multi-)matrix models:
\br
\pa^\ast_0 L = 0  \qquad \to \quad  \Sbr{M_{+}}{L} = - \one
\qquad ; \qquad
\pa^\ast_0 \Phi = 0  \qquad \to \quad  M_{+} \Phi = 0
\lab{01-constr-L}
\er
Eqs.\rf{01-constr-L}, using second eq.\rf{L-M},\rf{M-dress}
and first eq.\rf{paast} for $k=0$,
lead to the following constraints on $L$
\rf{f-5} and its DB-generating eigenfunction $\Phi$,
respectively:
\br
\sum_{l \geq 1} \frac{l+r}{r} t_{r+l} \partder{}{t_l} L +
\Sbr{t_1}{L}\, \d_{r,1} &= &- \one
\lab{L-constr} \\
\( \sum_{l \geq 1} \frac{l+r}{r} t_{r+l} \partder{}{t_l} + t_r \) \Phi &=&  0
\lab{P-constr}
\er

Recall now the formula \rf{tauok-1} for the $\t$-function of
the ${\sf cKP}_{r,m}$ hierarchy \rf{f-5}. Noticing that the
eigenfunctions $\Phi^{(k)}$ of the DB-transformed Lax operators $L^{(k)}$
satisfy the {\em same} constraint eq.\rf{P-constr} irrespective of the
DB-step $k$, we arrive at the following result (``string-equation''
constraint on the $\t$-functions) :
\begin{proposition}
The Wronskian $\t$-functions \rf{tauok-1} of ${\sf cKP}_{r,m}$
hierarchies \rf{f-5}, invariant under the lowest additional symmetry flow
\rf{01-constr-L},
satisfy the constraint equation:
\be
\( \sum_{l \geq 1} \frac{l+r}{r} t_{r+l} \partder{}{t_l} + nt_r \)
\frac{\t^{(n)}}{\t^{(0)}} = 0
\lab{tau-M-constr}
\ee
\label{proposition:taum}
\end{proposition}

As the simplest illustration of this proposition,
consider the discrete one-matrix model corresponding
to the generalized \BH hierarchy, {\sl i.e.},
to the chain of the Lax operators connected via DB transformations as
described in eqs.\rf{lkplus}-\rf{pkplus},
but with the {\em additional restriction} on $\p \equiv \Phi^{(0)}$
\rf{phi0} (coming from the orthogonal polynomial formalism) :
\be
\p =  \int d \l   \exp \(\sumi{r=1} t_r \l^r  \) \quad ,\;\;\; i.e. \;\;
\p^{(0)} (\l ) =1
\lab{P-0-constr}
\ee
The initial ``free'' eigenfunction \rf{P-0-constr} obeys the constraint
eq.\rf{P-constr} (for $r=1$) and, therefore,
proposition \ref{proposition:taum} (with $r=1$) yields precisely the
``string-equation'' in the one-matrix model:
\be
{\cal L}_{-1}^{(N)} W_N \lb \p, \pa \p, \ldots , \pa^{N-1} \p \rb =0
\qquad ,\qquad
{\cal L}_{-1}^{(N)} \equiv \sumi{k=2} k t_k \partder{}{t_{k-1}} + N t_1
\lab{string-eq-1MM}
\ee
Furthermore, as one can check directly \ct{oakpark}, the Wronskian
$\t$-function (second eq.\rf{phik}) with $\p$ restricted as in \rf{P-0-constr}
automatically satisfies all higher Virasoro constraints. Thus, we conclude
that for the particular class of \cKP hierarchies -- the generalized \BH
hierarchies \rf{lkplus}--\rf{phi0}, invariance under the lowest
additional-symmetry flow automatically triggers invariance under all higher
additional-symmetry flows as well.

\underline{Appendix: Technical Identities.}$\;\,$
We list here for convenience a number of
useful
technical identities, which have been used extensively throughout the text.

We work with calculus of pseudo-differential operators
based on the generalized Leibniz rule:
\be
D^n f  = \sumi{j=0} {n \choose j} (\pa^j f) D^ {n-j}
\lab{gleib}
\ee
For an arbitrary pseudo-differential operator $A$ we have the
following identity:
\be
\( \chi D \chi^{-1} A \chi D^{-1} \chi^{-1} \)_{+} =
\chi D \chi^{-1} A_{+}  \chi D^{-1} \chi^{-1}
-  \chi \pa_x \(\chi^{-1}  A_{+} (\chi ) \)  D^{-1} \chi^{-1}
\lab{aonchi}
\ee
where $A_{+}$ is the differential part of
$A= A_{+} + A_{-} = \sumi{i=0} A_i D^i + \sum_{-\infty}^{-1} A_i D^i$.

For a purely differential operator $K$ and arbitrary functions
$f,g$ we have an identity
\be
 \lb K\, , \, f D^{-1} g \rb_{-} =K (f) D^{-1} g-  f D^{-1} K^{\ast} (g)
\lab{tkppsi}
\ee
Another useful technical identity involves a product of two
pseudo-differential operators
of the form
$X_i = f_i D^{-1} g_i\;, \, i=1,2$ :
\be
X_1 X_2 = X_1 (f_2) D^{-1} g_2 + f_1 D^{-1} X^{\ast}_2 (g_1)
\lab{x1x2}
\ee
where $X_1 (f_2) = f_1 \pa_x^{-1} (g_1 f_2)$, etc.. .
{}From the above identity it follows the relation \ct{EOR95}:
\be
\(  L^k \)_{-} = \sum_{i=1}^m \sum_{j=0}^{k-1} L^{k-j-1} (\Phi_i) D^{-1}
\( L^{\ast}\)^{j} ( \Psi_i  )
\lab{lkminus}
\ee
for the \cKP Lax
operator \rf{f-5}.

Let us also list some useful identities involving {\DB}-like transformation
of pseudo-differential operators of the $X_i$-form above:
\br
T_a \( \Phi_a D^{-1} N \) T_a^{-1} &=& \( \Phi_a^2 N\) D^{-1} \Phi_a^{-1}
\lab{DB-like-1} \\
T_a \( M D^{-1} \Psi_a \) T_a^{-1} &=&
{\wti M} D^{-1} \( {\wti L}^\ast ({\wti \Psi_a})\) +
\lcurl T_a \( M \pa_x^{-1} (\Psi_a \Phi_a )\) \rcurl  D^{-1} \Phi_a^{-1}
\lab{DB-like-2} \\
T_a \( M D^{-1} N \) T_a^{-1} &=& {\wti M} D^{-1} {\wti N} +
\lcurl T_a \( M \pa_x^{-1} (N \Phi_a )\) \rcurl  D^{-1} \Phi_a^{-1}
\lab{DB-like-3} \\
\({\wti L}^{\ast}\)^{ s} ({\wti \Psi_a}) &=&
- \Phi_a^{-1} \pa_x^{-1} \( \Phi_a \({L^\ast}\)^{s-1} (\Psi_a )\)
\lab{DB-like-4}
\er
where $\Phi_a$ is one of the eigenfunctions of a \cKP Lax operator $L$
\rf{f-5} and
\br
T_a &\equiv& \Phi_a D \Phi_a^{-1} \qquad ,\qquad  {\wti \Psi_a} = \Phi_a^{-1}
\nonu \\
{\wti M} &\equiv& T_a (M) = \Phi_a \pa_x \( \Phi_a^{-1} M\) \quad ,\quad
{\wti N} \equiv {T_a^{-1}}^\ast (N) = - \Phi_a^{-1} \pa_x^{-1} \( \Phi_a N\)
\nonu
\er

Finally, let us recall the following important composition formula for
Wronskians \ct{Wronski} :
\be
T_k \, T_{k-1}\, \cdots\, T_1 (f ) \; =\; { W_{k} (f) \over W_k}
\lab{iw}
\ee
where
\br
T_j = { W_{j} \over W_{j-1} } D { W_{j-1} \over W_{j} } =
\( D + \( \ln { W_{j-1} \over W_{j} } \)^{\pr} \) \quad;\quad W_{0}=1
\lab{transf} \\
W_k \equiv W_k \lb \psi_1, \ldots ,\psi_k \rb =
\det {\Bigl\Vert} \pa_x^{i-1} \psi_j {\Bigr\Vert}
\quad ,\quad
W_{k-1} \(f \)\equiv W_{k} \lb \psi_1, \ldots ,\psi_{k-1}, f\rb
\lab{W-def}
\er

\end{document}